\documentclass[a4paper,UKenglish]{lipics}
\usepackage{microtype}
\usepackage{amsmath,amssymb,amstext}
\usepackage{enumerate}
\usepackage{proof}
\usepackage{wrapfig}
\usepackage{framed}
\usepackage{tikz}
\usetikzlibrary{arrows,automata,backgrounds,calc,chains,fadings,folding,decorations.fractals,fit,patterns,positioning,mindmap,shadows,shapes.geometric,shapes.symbols,through,trees,decorations.markings,decorations.text}
\colorlet{dblue}{blue!40!black}

\newcommand{\tuple}[1]{\langle\,#1\,\rangle}

\newcommand{\ter}[2]{T(#1,#2)}
\newcommand{\vars}{\mathcal{X}}

\newcommand{\pow}[1]{\mathcal{P}(#1)}
\newcommand{\inter}[3]{[#1,#2]_{#3}}
\newcommand{\interg}[2]{[#1]_{#2}}
\newcommand{\lang}[1]{\mathcal{L}(#1)}

\renewcommand{\emptyset}{\varnothing}

\newcommand{\sap}{\mathit{ap}}
\newcommand{\ap}[2]{\sap(#1,#2)}
\newcommand{\combS}{\mathit{S}}

\title{Non-termination using Regular Languages\footnote{Published at the International Workshop on Termination 2014.} ---~International Workshop on Termination ---}
\titlerunning{Non-termination using Regular Languages}
                             
\author[1]{J\"{o}rg~Endrullis}
\author[2]{Hans~Zantema}

\affil[1]{VU University Amsterdam, The Netherlands}
\affil[2]{Eindhoven University of Technology, The Netherlands}  

\authorrunning{J. Endrullis and H. Zantema}

\Copyright{J. Endrullis and H. Zantema}


\begin{document}

\maketitle

\begin{abstract}
  We describe a method for proving non-termination 
  of term rewriting systems that do not admit looping reductions.
  As certificates of non-termination, we employ regular (tree) automata.
\end{abstract}

\section{Introduction}

We describe a method for proving non-termination 
of term rewriting systems that do not admit looping reductions, that is, 
reductions from a term $t$ to a term $C[t\sigma]$ containing a substitution instance of $t$.
For this purpose, we employ tree automata as certificates of non-termination.
For proving non-termination of a term rewriting system $R$,
we search a tree automaton $A$ whose language $\lang{A}$
is not empty, weakly closed under rewriting and every term of the language contains a redex occurrence.
We have fully automated the search for these certificates employing SAT-solvers.

All automata that we use as example in this paper
have been found automatically;
this concerns in particular fully automated proofs of non-termination for the following two rewrite systems.

\begin{example}\label{ex:lr}
  We consider the following string rewriting system:
  \begin{align*}
   zL &\to Lz & 
   Rz &\to zR &
   bL &\to bR &
   Rb &\to Lzb
  \end{align*}
  This rewrite system admits no reductions of the form $s \to^* \ell s r$.
\end{example}

\begin{example}\label{ex:S}
  We consider the $S$-rule from combinatory logic:
  \begin{align*}
    \ap{\ap{\ap{\combS}{x}}{y}}{z} \to \ap{\ap{x}{z}}{\ap{y}{z}}
  \end{align*}  
  For the $S$-rule it is known that there are no reductions $t \to^* C[t]$ for ground terms $t$, see~\cite{wald:2000}.
  For open terms $t$ the existence of reductions $t \to^* C[t\sigma]$ is open.
\end{example}

It turns out that the method can be fruitfully applied 
to obtain non-termination proofs of several string rewriting systems
that have remained unsolved in the last full run of the termination competition.
% \smallskip
% 
% {\small
% \begin{itemize}
%   \item \verb=SRS_Standard/Trafo_06/un16.trs=
%   \item \verb=SRS_Standard/Secret_07_SRS/num-520.trs= -- Matchbox 2007
%   \item \verb=SRS_Standard/Waldmann_07_size12/size-12-alpha-2-num-12.trs= -- Matchbox 2007
%   \item \verb=SRS_Standard/Waldmann_07_size12/size-12-alpha-2-num-17.trs= -- Matchbox 2007
%   \item \verb=SRS_Standard/Waldmann_07_size12/size-12-alpha-2-num-18.trs= -- Matchbox 2007
%   \item \verb=SRS_Standard/Waldmann_07_size12/size-12-alpha-3-num-120.trs= -- Matchbox 2007
%   \item \verb=SRS_Standard/Waldmann_07_size12/size-12-alpha-3-num-211.trs=
%   \item \verb=SRS_Standard/Waldmann_07_size12/size-12-alpha-3-num-224.trs=
%   \item \verb=SRS_Standard/Waldmann_07_size12/size-12-alpha-3-num-355.trs=
%   \item \verb=SRS_Standard/Waldmann_07_size12/size-12-alpha-3-num-384.trs=
%   \item \verb=SRS_Standard/Waldmann_07_size12/size-12-alpha-3-num-385.trs=
%   \item \verb=SRS_Standard/Waldmann_07_size12/size-12-alpha-3-num-540.trs= -- Matchbox 2007
% \end{itemize}
% }

\paragraph*{Related Work}

The paper~\cite{gese:zant:1999} investigates necessary conditions for the existence of loops.
The work~\cite{zankl:2007} employs SAT solvers to find loops, 
\cite{zank:2010} uses forward closures to find loops efficiently, and
the wook \cite{cloops} introduces `compressed loops' to 
find certain forms of (possibly very long) loops. 

Non-termination beyond loops has been investigated in~\cite{oppe:2008} and~\cite{emme:enge:gies:2012};
we note that Example~\ref{ex:S} cannot be handled by these techniques.

Here we prove non-looping non-termination on regular languages.
The converse, local termination on regular languages, has been investigated in~\cite{endr:vrij:wald:2010}.
Regular (tree) automata have been fruitfully applied to a wide rage of properties of term rewriting systems: 
for proving termination~\cite{matchbounds:2007,matchbounds:2006,matchbounds:2009},
for infinitary normalization~\cite{endr:grab:hend:klop:vrij:2009},
for proving liveness~\cite{liveness:2010},
and for analysing reachability and deciding the existance of common reducts~\cite{lata:2014,endr:grab:klop:oost:2011}.

\section{Non-termination and Weakly Closed Languages}

\begin{definition}
  Let $L \subseteq \ter{\Sigma}{\emptyset}$ a language and $R$ a TRS over $\Sigma$.
  Then $L$ is called:
  \begin{itemize}
    \item \emph{closed} under rewriting 
      if for every $t \in L$ and $s$ such that $t \to s$, one has $s \in L$, and
    \item \emph{weakly closed} under rewriting
      if for every $t \in L$ that is not in normal form, there exists $s \in L$ such that $t \to_R s$.
  \end{itemize}
\end{definition}
\smallskip

The following theorem describes the basic idea that we employ for proving non-termination.

\begin{theorem}\label{thm:main}
  A term rewriting system $R$ over $\Sigma$ is non-terminating 
  if and only if there exists a non-empty language $L \subseteq \ter{\Sigma}{\vars}$
  such that
  \begin{enumerate}[(i)]
    \item every $t\in L$ contains a redex (that is, $t\to s$ for some term $s$), and
    \item $L$ is weakly closed under rewriting. \qed
  \end{enumerate}
\end{theorem}
A language fulfilling the properties of Theorem~\ref{thm:main} 
is also called a \emph{recurrence set}, see~\cite{cook}.

To automate this method, we need to restrict to a certain family of languages.
In this paper, we consider regular tree languages.
To guarantee that the language of a tree automaton is weakly closed under rewriting,
we check that the language is not empty and that the automaton is a quasi-model (see Definition~\ref{def:quasi}) for the rewrite system.
The latter condition is actually too strict;  
it implies that the languages is not only weakly closed, but also closed under rewriting.
In future, we plan to relieve this restriction.

\section{Tree Automata}

\begin{definition}\label{def:nfa}
  A \emph{(nondeterministic finite) tree automaton $A$ over a signature $\Sigma$} 
  is a tuple $A = \tuple{Q,\Sigma,F,\delta}$
  where 
  \begin{enumerate}[(i)]
    \item $Q$ is a finite set of \emph{states},
    \item $F \subseteq Q$ is a set of \emph{accepting states}, and
    \item $\{\delta_f\}_{f\in\Sigma}$ is a family of \emph{transition relations} such that
      for every $f\in \Sigma$:
      \begin{align*}
        \delta_f \subseteq Q^n \times Q
      \end{align*}
      where $n$ is the arity of $f$. 
  \end{enumerate}
% 
%   \noindent
%   The tree automaton $A$ is called \emph{deterministic} if \ldots
\end{definition}

In examples, we often write the transition relation $\delta_f$ as $\to_f$.
\begin{example}\label{ex:automaton:lr}
  The following is a tree automaton for the signature in Example~\ref{ex:lr}.
  We consider string rewriting systems as term rewriting systems
  by interpreting all symbols as unary and adding a special constant $\varepsilon$
  to denote the end of the word.
  Let $A_{LR} = \tuple{Q,\Sigma,F,\to}$ where $Q = \{0,1,2,3\}$, $\Sigma = \{b,L,R,0,\varepsilon\}$, $F = \{3\}$
  and 
  \begin{align*}
    &\to_{\varepsilon} 0 & 1 &\to_{z} 1 & 0 &\to_b 1 & 1 &\to_R 2 & 1 &\to_L 2\\
    && 2 &\to_{z} 2 & 2 &\to_b 3 
  \end{align*}
  The transition relation for $\varepsilon$ can be thought of as defining the initial states (here $0$)
  of a word automaton.
\end{example}

\begin{example}\label{ex:automaton:s}
  The following is a tree automaton for Example~\ref{ex:S}.
  Let $A_{S} = \tuple{Q,\Sigma,F,\to}$ where $Q = \{0,1,2,3,4\}$, $\Sigma = \{\sap,\combS\}$, $F = \{4\}$
  and
  \begin{align*}
    &\to_{\combS} 0 & 
    (0,0) &\to_{\sap} 1 &
    (1,0) &\to_{\sap} 2 & 
    (2,2) &\to_{\sap} 3 & 
    (3,3) &\to_{\sap} 3 
    \\ 
    && (0,2) &\to_{\sap} 2 &
    && (2,3) &\to_{\sap} 3 & 
    (3,3) &\to_{\sap} 4
    \\ 
    && (0,3) &\to_{\sap} 2
  \end{align*}
  In Example~\ref{ex:s:accept} we show that this automaton accepts the term $\combS\combS\combS(\combS\combS\combS)(\combS\combS\combS(\combS\combS\combS))$.
\end{example}

% \begin{tabular}{c|ccccccccccccccccccccccccc}
%  x0 & 0 & 0 & 0 & 0 & 0 & 1 & 1 & 1 & 1 & 1 & 2 & 2 & 2 & 2 & 2 & 3 & 3 & 3 & 3 & 3 & 4 & 4 & 4 & 4 & 4 \\
%  x1 & 0 & 1 & 2 & 3 & 4 & 0 & 1 & 2 & 3 & 4 & 0 & 1 & 2 & 3 & 4 & 0 & 1 & 2 & 3 & 4 & 0 & 1 & 2 & 3 & 4 \\
%  \hline
%    &   &   &   &   &   &   &   &   &   &   &   &   &   &   &   &   &   &   &   &   &   &   &   &   &   \\
%  1 & 1 &   &   &   &   &   &   &   &   &   &   &   &   &   &   &   &   &   &   &   &   &   &   &   &   \\
%  2 &   &   & 1 & 1 &   & 1 &   &   &   &   &   &   &   &   &   &   &   &   &   &   &   &   &   &   &   \\
%  3 &   &   &   &   &   &   &   &   &   &   &   &   & 1 & 1 &   &   &   &   & 1 &   &   &   &   &   &   \\
%  4 &   &   &   &   &   &   &   &   &   &   &   &   &   &   &   &   &   &   & 1 &   &   &   &   &   &   \\
% \end{tabular}

\begin{definition}\label{def:interpretation}
  Let $A = \tuple{Q,\Sigma,F,\delta}$ be a tree automaton over $\Sigma$.
  For terms $t \in \ter{\Sigma}{\vars}$ and assignments $\alpha : \vars \to \pow{Q}$
  we define the \emph{interpretation} $\inter{t}{\alpha}{A}$ by:
  \begin{align*}
    \inter{x}{\alpha}{A} &= \alpha(x) \\
    \inter{f(t_1,\ldots,t_n)}{\alpha}{A} &= \{ q \mid 
      (q_1,\ldots,q_n) \in \inter{t_1}{\alpha}{A} \times\ldots\times \inter{t_n}{\alpha}{A},\;
      \tuple{(q_1,\ldots,q_n),q} \in \delta_f \}
  \end{align*}
  Whenever $A$ is clear from the context, we write $\inter{t}{\alpha}{}$ as shorthand for $\inter{t}{\alpha}{A}$.
  For ground terms $t$, the interpretation is independent of $\alpha$, 
  allowing is to write $\interg{t}{A}$ or $\interg{t}{}$ for short.
\end{definition}

\begin{example}\label{ex:s:interpretation}
  We use the automaton $A_S$ from Example~\ref{ex:automaton:s}.
  Let $\alpha(x) = \{2\}$, then we have:
  \begin{gather*}
    \inter{\combS}{\alpha}{} = \{0\} \quad\quad\quad
    \inter{\ap{\combS}{\combS}}{\alpha}{} = \{1\} \quad\quad\quad
    \inter{\ap{\ap{\combS}{\combS}}{\combS}}{\alpha}{} = \{2\} \\
    \inter{\ap{x}{x}}{\alpha}{} = \{3\} \quad\quad\quad
    \inter{\ap{\ap{x}{x}}{\ap{x}{x}}}{\alpha}{} = \{3,4\} \quad\quad\quad
  \end{gather*}
\end{example}

\begin{definition}
  Let $A = \tuple{Q,\Sigma,F,\delta}$ be a tree automaton over $\Sigma$.
  The \emph{language $\lang{A}$ accepted by $A$} is the set
  $\lang{A} = \{ t \mid t \in \ter{\Sigma}{\emptyset},\; \interg{t}{A} \cap F \ne \emptyset\}$
  of ground terms.
\end{definition}

\begin{example}
  The automaton in Example~\ref{ex:automaton:lr} accepts all words of the form $b\;z^*\;(L|R)\;z^*\;b$,
  that is,
  all words that start with $b$, end with $b$, contain one $L$ or $R$
  and otherwise only $z$.
\end{example}

\begin{example}\label{ex:s:accept}
  We continue Example~\ref{ex:s:interpretation}:
  \begin{gather*}
    \interg{\ap{\ap{\combS}{\combS}}{\combS}}{} = \{2\} 
    \quad\quad\quad
    \interg{ \ap{ \;\ap{\ap{\combS}{\combS}}{\combS} \;}{\; \ap{\ap{\combS}{\combS}}{\combS} \;} }{} = \{3\} 
    \\
    \interg{ \ap{ \;\ap{ \ap{\ap{\combS}{\combS}}{\combS} }{ \ap{\ap{\combS}{\combS}}{\combS} } \;}{ \;\ap{ \ap{\ap{\combS}{\combS}}{\combS} }{ \ap{\ap{\combS}{\combS}}{\combS} } \;} }{} = \{3,4\} 
  \end{gather*}
  Thus $F \cap \interg{\combS\combS\combS(\combS\combS\combS)(\combS\combS\combS(\combS\combS\combS))}{} = \{4\} \ne \varnothing$
  and hence the term is accepted by the automaton.
\end{example}

\section{Closure under Rewriting}

\begin{definition}\label{def:quasi}
  A tree automaton $A = \tuple{Q,\Sigma,F,\delta}$ is a \emph{quasi-model}
  for a term rewriting system $R$ over $\Sigma$ if
  $\inter{\ell}{\alpha}{A} \subseteq \inter{r}{\alpha}{A}$
  for every $\ell \to r \in R$ and $\alpha : \vars \to \pow{Q}$.
\end{definition}

Actually, it suffices to check the property
$\inter{\ell}{\alpha}{A} \subseteq \inter{r}{\alpha}{A}$
for assignments $\alpha : \vars \to \pow{Q}$ that map variables to singleton sets.

\begin{lemma}\label{lem:quasi}
  A tree automaton $A = \tuple{Q,\Sigma,F,\delta}$ is a \emph{quasi-model}
  for a term rewriting system $R$ over $\Sigma$ iff
  $\inter{\ell}{\alpha}{A} \subseteq \inter{r}{\alpha}{A}$
  for every $\ell \to r \in R$ and $\alpha : \vars \to \{\{q\}\mid q\in Q\}$.
\end{lemma}

\begin{example}
  It is not difficult to check that the automaton $A_{LR}$ from Example~\ref{ex:automaton:lr}
  is a quasi-model for rewrite system in Example~\ref{ex:lr}.
\end{example}

\begin{example}
  We consider the automaton $A_S$ from Example~\ref{ex:automaton:s}.
  We write $(a,b,c) \to d$ if $d \in \inter{\ell}{\alpha}{}$ 
  when $\alpha(x) = \{a\}$, $\alpha(y) = \{b\}$, $\alpha(z) = \{c\}$.
  Then for $\inter{\ell}{\alpha}{}$ we have:
  \begin{align*}
    (0,0,2) &\to 1 &
    (2,2,3) &\to 3 &
    (2,3,3) &\to 3 &
    (3,2,3) &\to 3 &
    (3,3,3) &\to 3 \\
    (0,0,3) &\to 1 &
    (2,2,3) &\to 4 &
    (2,3,3) &\to 4 &
    (3,2,3) &\to 4 &
    (3,3,3) &\to 4
  \end{align*}
  The interpretation $\inter{r}{\alpha}{}$ has all the above and additionally:
  \begin{align*}
    (0,2,2) &\to 3 & (1,1,0) &\to 3 & (2,2,2) &\to 3 \\
    (0,2,3) &\to 3 & && (2,2,2) &\to 4 \\
    (0,3,3) &\to 3 
  \end{align*}
  As a consequence $A_S$ is a quasi-model for the $S$-rule.
\end{example}

The following theorem is immediate:
\begin{theorem}
  Let $A = \tuple{Q,\Sigma,F,\delta}$ be a tree automaton 
  and $R$ a term rewriting system over $\Sigma$.
  If $A$ is a quasi-model for $R$ then the language of $A$ is closed under rewriting.
  \qed
\end{theorem}

\section{Ensuring Redex Occurrences}

Next, we want to guarantee that every term in the language $\lang{A}$ of an automaton $A$ 
contains a redex with respect to the term rewriting system $R$.
For left-linear systems $R$, 
this problem can be reduced to deciding the inclusion of regular languages.

Let $R$ be a left-linear term rewriting system.
Then the set of ground terms containing a redex is a regular tree language.
A deterministic automaton $B$ for this language can be constructed
using the overlap-closure of subterms of left-hand sides, see further~\cite{endr:hend:2010,endr:hend:2011}.

\begin{example}\label{ex:redex:s}
  The following tree automaton $C = \tuple{Q,\Sigma,F,\to}$ accepts the language of
  ground terms that contain a redex occurrence with respect to the $S$-rule.
  Here $Q = \{0,1,2,3\}$, $\Sigma = \{\sap,\combS\}$, $F = \{3\}$
  and
  \begin{align*}
    &\to_{\combS} 0 & 
    (0,q) &\to_{\sap} 1 &
    (1,q) &\to_{\sap} 2 & 
    (2,q) &\to_{\sap} 3 &
    (3,q) &\to_{\sap} 3 &
    (q,3) &\to_{\sap} 3 
  \end{align*}
  for all $q \in \{0,1,2\}$.
\end{example}

As a consequence the problem of checking whether every term in $\lang{A}$ contains a redex
boils down to checking that $\lang{A} \subseteq \lang{B}$.
For non-deterministic $A$ and deterministic $B$, this property can be decided by
constructing the product automaton and considering the reachable states.

\begin{definition}
  The \emph{product $A\cdot B$} of tree automata $A = \tuple{Q,\Sigma,F,\delta}$ and  $B = \tuple{Q',\Sigma,F',\delta'}$
  is the automaton $C = \tuple{Q \times Q',\Sigma,\emptyset,\gamma}$ where
  for every $f \in \Sigma$ of arity $n$,
  we define the transition relation $\gamma_f \subseteq (Q \times Q')^n \times (Q \times Q')$ by
  \begin{align*}
    \tuple{(q_1,p_1),\ldots,(q_n,p_n)} \mathrel{\gamma} (q',p') 
    \;\iff\;
    \tuple{q_1,\ldots,q_n} \mathrel{\delta_{f}} q' 
    \;\wedge\;
    \tuple{p_1,\ldots,p_n} \mathrel{\delta'_{f}} p' 
  \end{align*}
\end{definition}

\begin{definition}
  The set of \emph{reachable states} of a tree automaton $A = \tuple{Q,\Sigma,F,\delta}$
  is the smallest set $S \subseteq Q$ such that
  $q \in S$ whenever $\tuple{q_1,\ldots,q_n} \mathrel{\delta_f} q$ 
  for some $q_1,\ldots,q_n \in S$ and $f\in\Sigma$ with arity $n$.
\end{definition}

The following theorem gives a method for checking $\lang{A} \subseteq \lang{B}$
without the need for determinising $A$ (only $B$ needs to be deterministic).

\begin{theorem}\label{thm:redex}
  Let $A = \tuple{Q,\Sigma,F,\delta}$ and $B = \tuple{Q',\Sigma,F',\delta'}$ be tree automata
  such that $B$ is deterministic.
  Let $S$ be the set of reachable states of the product automaton $A\cdot B$.
  Then $\lang{A} \subseteq \lang{B}$ if and only if for all $(q,p) \in S$
  it holds that $q \in F \implies p \in F'$.
\end{theorem}

\begin{example}
  The reachable states of product automaton $A_S \cdot C$ 
  of the automata $A_S$ from Example~\ref{ex:automaton:s}
  and $C$ from Example~\ref{ex:redex:s} are
  $(0,0), (1,1), (2,2), (2,1), (3,3), (3,2), (2,3), (4,3)$.
  The only state $(q,q')$ such that $q$ is accepting in $A_S$ is $(4,3)$ and
  $3$ is an accepting state of $C$.
  Thus the conditions of Theorem~\ref{thm:redex} are fulfilled and hence $\lang{A_S} \subseteq \lang{C}$.
  Thus every term accepted by $A_S$ contains a redex.
\end{example}

\section{Future Work}

We plan to investigate whether the method described in this paper can be
fruitfully extended from regular automata to pushdown automata, that is, context-free languages.
For this purpose, it is important that 
it is decidable whether a context-free language is a subset of a regular language 
(the language of terms containing left-linear redex occurrences).
However, it remains to be investigated whether context-free certificates
can be found efficiently.

\bibliography{main.bib}

\end{document}